# AI in Music and Sound: Pedagogical Reflections, Post-Structuralist Approaches and Creative Outcomes in Course Practice

**Guilherme Coelho**
Audio Communication Department
Technische Universität Berlin
*guilherme.coelho@thenewcentre.org*

## Abstract

This paper presents a pedagogical and conceptual account of the course AI in Music and Sound: Modalities, Tools, and Creative Applications, offered within the Music Informatics and Media Art module of an M.Sc. in Audio Communication. The course engaged students with a range of AI modalities such as symbolic composition, voice synthesis, timbre transfer, neural audio synthesis, and text-to-audio systems, combining theoretical reflection with practice-based experimentation. Its central pedagogical move is a paired-études design: each modality is approached first through its intended affordances and then through a deliberately reframed or "misused" exercise that surfaces representational limits and alternative behaviours. Framed by medium theory and post-structuralist inquiry, we treated AI as a transmodal conduit—a system that translates and perturbs musical signs across textual, symbolic, timbral, and audio domains. Evidence from student work and reflections indicates growth in technical fluency, medium awareness, and critical literacy, alongside the cultivation of experimental method and process-oriented listening. The paper outlines the course architecture, assessment design, and representative projects, and distils a set of design patterns for AI-music pedagogy (e.g., semantic destabilisation in text-to-audio; latent-space materialism in timbre transfer; prompt-conditioned interplays). It concludes with pedagogical recommendations that integrate creative practice with medium awareness and with cultural–epistemic analysis of AI technologies, preparing students to participate in how AI is understood, developed, and deployed within creative communities.

## 1. Introduction

The accelerating integration of artificial intelligence into music-making practices presents new tools for composition and production alongside complex conceptual challenges regarding the nature of sound, musicianship, and creativity. As AI systems increasingly participate in the generation and transformation of musical material, pedagogical approaches should respond by developing technical fluency while exploring and interrogating the epistemological and aesthetic paradigms that underwrite these technologies. This paper advances a practice-based, theory-attentive pedagogy organised around a *paired-études* model and a conception of AI as a transmodal conduit, showing how students develop medium awareness, experimental method, and critical literacy across symbolic, timbral, vocal, and text-to-audio modalities.

Over the past year, I have designed and taught graduate-level courses examining the intersections of AI, creative practice, and musical discourse. One such course, titled *Machine Learning and Creativity in Music and Sound*, explored the tension between AI and human creativity, examining how music functions as an established language system with its own syntactic and semantic structures. This course considered training data as a creative milieu, drawing on structuralist and post-structuralist frameworks, Deleuzian concepts of territorialization, intertextuality (Kristeva, 1980), meta-narratives (Lyotard, 1984) and academic definitions of creativity to consider our dispositions toward musical meaning and the sociolinguistic dimensions of sound. In contrast, the more recent course *AI in Music and Sound: Modalities, Tools, and Creative* Applications extends that inquiry into a highly practice-based format, introducing multiple AI modalities through artistic exercises and hands-on explorations that test and reframe the systems' affordances.

Offered within the Music Informatics and Media Art module of the Audio Communication M.Sc. program of TU Berlin and open to students from other programs, this latter course—the focus of this paper—aimed to cultivate critical reflection and experimental practice. The course had students from diverse disciplinary backgrounds and musical experience levels, including students who play instruments in bands, classically trained musicians, in-the-box producers, novices who have experimented with DAWs, and students with zero music-making experience. Rather than positioning AI as a neutral compositional assistant, the course approached it as a transmodal conduit: a system capable of mediating, recontextualizing, and destabilizing musical signs across symbolic, timbral, textual, and audio domains enabling cross-material[1] translation rather than mere stylistic imitation. Framed by post-structuralist theory, new materialist philosophies, and aesthetic inferentialism,[2] the course emphasized music's condition as a medium—a historically and technologically embedded apparatus of meaning production, representation, and cultural inscription. This awareness of medium functioned as a theoretical anchor across the course's practical and conceptual dimensions, inviting students to think about the interactions of AI systems and the infrastructures, histories, and codes they traverse.

There is a marked scarcity of pedagogical models that integrate AI into sound-based creative practices while remaining attentive to the conceptual and critical complexities of these technologies. Despite the proliferation of AI tools in creative domains, few academic offerings explore these technologies from both theoretical and practice-based orientations, and even fewer consider AI as a medium in itself—capable of altering how music is conceived, encoded, and situated. Many academic programs tend to frame AI narrowly, either through the lens of technical development (focusing on algorithm design and implementation) or as an established production aid (reinforcing established representational logics and existing conventions). Seeking to depart from established instruction, the course treated AI systems as unstable yet generative sites of translation and indeterminacy. Through modalities such as timbre transfer and text-to-audio, students explored how these systems traverse textual, timbral, and conceptual regimes, rerouting musical signs across media. Framed in this way, AI was taught as a medium in its own right—one in which meaning is at once reproduced and reconfigured, and in which "musicality" becomes an object of inquiry rather than an assumption. Consequently, the course fostered medium awareness, treating music as a stratified cultural–technological apparatus and foregrounding latent space as a locus of sonic emergence.

This paper proceeds in five parts. Section 2 lays out the pedagogical framework of the course, focusing on its philosophical underpinnings, learner-centered approach, and commitment to conceptual experimentation. Section 3 outlines the course structure and content, providing an overview of the weekly sessions and highlighting key modalities and topics. Section 4 discusses student assignments and project work, illustrating how the dual-étude structure fostered engagement with established affordances and critical departures from them. Section 5 synthesises student learning and articulates implications for a critical AI pedagogy. The Conclusion consolidates the paper's contributions and sketches extensions of this framework across interdisciplinary and cross-cultural contexts, highlighting practices that counter homogenisation while leveraging AI's distinctive affordances.

---

[1] Cross-materialism: compositional practice that traverses distinct material substrates (e.g., timbre transfer of singing to violin), foregrounding entanglement and transformation over resemblance.

[2] Aesthetic inferentialism: drawing on inferentialist philosophy, the view that musical meaning arises from the practical relations and commitments enacted in making and listening, not from fixed denotations.

## 2. Pedagogical Framework

### 2.1 Theoretical Foundations and Philosophical Underpinnings

At the core of this course's design lies a commitment to post-structuralist, media-aware, and deconstructive pedagogical strategies[3] that seek to open conceptual and creative spaces for the interrogation, expansion, and recontextualization of musical form, rather than reifying fixed notions. The course foregrounded music fundamentally as a medium, expanding on the notion introduced earlier to emphasize its function as a culturally and historically situated system of signs, affordances, and representations, while drawing from its understanding as organized sound (Schaeffer, 1966), inscription between noise and silence (Attali, 1985), and phenomenological experience (Cage, 1961). Through the lens of medium theory, students were encouraged to interrogate how music comes to be legible as such through particular infrastructures of notation, reproduction, perception, and now, algorithmic mediation, alongside examining what music *is* or *does*. Instead of prescribing a fixed understanding of what music should sound like or how it should be made, the course encouraged students toward the conditions and processes through which music is constituted and represented, and what music could become when mediated through the logics, constraints, and affordances of AI systems.

This orientation is complemented by an account of emergence theory, drawing on Deleuzian philosophy, in which creative processes arise from the dynamic interplay of heterogeneous elements rather than from predetermined structures (Deleuze & Guattari, 1987). This proved particularly apt for analysing how AI systems yield novel sonic formations through the entanglement of training data, model architectures, and user interventions: outcomes are not simple input–output functions but the result of relational dynamics among multiple agents and materials. For instance, repeated generations with identical prompts in text-to-audio systems (e.g., *Udio*) produce non-identical outputs; moreover, variance is increased as prompts became more semantically open (e.g., prompt: "plunderphonics, deconstructed human music"). This non-deterministic dispersion—driven by stochastic sampling and latent conditioning—illustrates emergent behaviour: outputs are not simple mappings of inputs but contingent articulations produced by the interplay of datasets, model dynamics, and user interventions.

Within this paradigm, AI is approached as a transmodal conduit—a system that traverses, entangles, and re-routes textual, sonic, and conceptual codes across modalities—rather than as a neutral assistant for automating composition.[4] This framing was deeply informed by New Materialist

---

[3] The terms *post-structuralist*, *media-aware*, and *deconstructive* are used here to describe a pedagogical orientation that foregrounds the instability of meaning, the constructedness of musical categories, and the role of mediation in shaping artistic production. Post-structuralist theory (drawing from figures such as Derrida, Foucault, and Deleuze) was mobilized to encourage students to interrogate systems of representation and to treat AI not as a fixed generator of form but as a site of différance, emergence, and discontinuity. A *media-aware* approach emphasized that music is always mediated—through notation, instruments, interfaces, and now algorithms—each of which carries specific affordances and cultural inscriptions. *Deconstructive* practices were encouraged to reveal the tensions, limits, and latent potentials within AI systems, allowing students to explore not only what AI can do, but how it frames, fragments, and recontextualizes sonic meaning.

[4] I use the term *transmodal conduit* to denote a system that mediates translation, transformation, and recombination of information across distinct semiotic and material domains (e.g., from text to sound). Such a conduit enables hybrid forms of representation by allowing conceptual and aesthetic materials to migrate

perspectives (Barad, 2007; Bennett, 2010) and Aesthetic Inferentialism (Brandom, 2000), positioning AI as an active co-articulator in the construction, mediation, and reconfiguration of musical representations and forms. In this context, AI is understood to engage in semantic reconstruction and material recomposition through complex, situated processes of inference, representation, and emergence.

A central element was the semiotic distinction between *signifier* and *signified* (Saussure, 1916/1983), and its destabilisation by post-structuralist thought. Structuralist approaches conceive of the sign as a relatively stable relationship between perceptible form and conceptual content. Post-structuralist perspectives, particularly Derrida's (1976) notion of *différance*, emphasises instability, deferral, and the contextual contingency of this relation. In musical terms, this reframing shifts attention from sound as a fixed bearer of meaning toward sound as a relational construct, shaped by cultural codes, listening practices, and technologies. This perspective aligned with Foucault's understanding of discourse (1972) as a system that produces knowledge and meaning rather than simply reflecting pre-existing truths, positioning AI not as neutral repositories but as culturally inscribed archives that condition generative possibilities.

AI further complicates representation through latent encodings: high-dimensional vectors abstracted from surface features during learning. While a structuralist reading treats these vectors as compressed models of regularities, a post-structuralist reading treats them as sites of ongoing mediation and translation in which any putative "signified" is reconstructed via probabilistic inference and contextual activation. Pedagogically, the autoencoder paradigm —a model that compresses data into a latent space and then reconstitutes it—as a central metaphor for exploring both the potentials and indeterminacies of AI mediation. Latent space was approached as a dynamic and unstable field in which meaning arises through activation, recombination, and material transformation. In this framing, musical meaning emerges through ongoing processes of signification, with the relation between input and output mediated by opaque, contingent, and mutable representational layers.

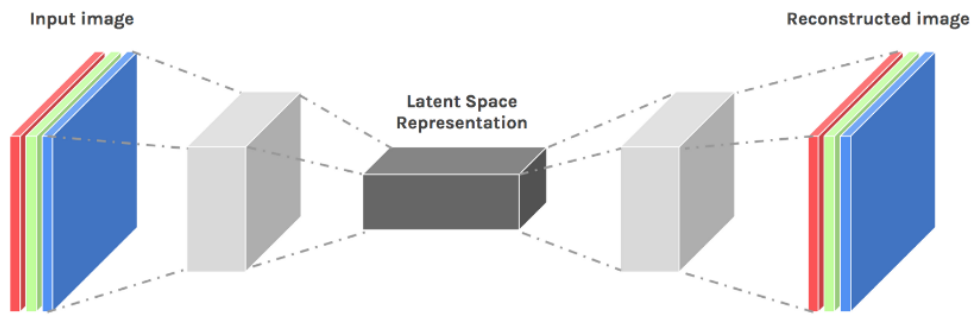

**Figure 1.** Autoencoder paradigm as a pedagogical metaphor for AI mediation.

Schematic of an encoder (left) compressing an input signal (image, audio, or symbolic sequence) into a latent representation **z**, and a decoder (right) reconstructing an output from **z**. In the course's framing, the latent space is not a neutral bottleneck but a site of mediation: statistical priors and architectural constraints translate, filter, and recombine features so that any "signified" is reconstructed rather than retrieved. Differences between input and output index representational bias and material transformation. We use this diagram to discuss timbre transfer, text-to-audio, and

---

between media rather than remain confined to a single modality. In this course the term frames AI as an *intermediary medium* that reshapes meaning across representational systems, not as a unidirectional production tool.

neural synthesis as practices of activating and traversing z—where meaning arises through activation, recombination, and contingent mappings across modalities.

Deleuze and Guattari's concepts of territorialisation, deterritorialisation, and reterritorialisation (Deleuze & Guattari, 1987) provided further tools for understanding how AI systems can encode, displace, and reconfigure sonic materials. Students were encouraged to treat musical data as perpetually mutable, undergoing transformations across symbolic, timbral, and affective registers. This perspective was complemented by the concept of *cross-materialism*[5], used to describe traversals between materially distinct yet possibly adjacent forms, such as converting environmental noise into rhythmic structures, screams into violins, or ambient noise into synthetic vocality.

Throughout the course, students were encouraged to attend to the ways in which their own interventions—through input audio, prompt design, aesthetic decisions, critical framing, or conceptual curation—participated in the shaping of the AI system's outputs. The notion that these systems bore traces of both human intention and machinic idiosyncrasy was central to our approach. Rather than seeking to "master" the tools in pursuit of stylistic fluency, students were prompted to explore them as systems of behavior, capable of both reflecting and distorting their inputs in unexpected ways.

An early exercise on voice synthesis illustrates this stance. Students employed pre-trained singing-voice models to render diverse sources via timbre transfer. The resulting artefacts often exhibited ontological ambiguity: familiar vocal timbres unsettled by computational artefacts, non-standard phrasing, spectral discontinuities, and flattened affective contours. Pedagogical attention was directed toward these moments of *semiotic slippage*[6], where language faltered, articulation dissolved, or expressive residues surfaced unintentionally. We treated such moments of as aesthetic events in their own right—sites where meaning and form emerge through friction, divergence, and surplus—thereby foregrounding the productive instability of AI. This dynamic was especially acute in *text-to-audio*, where words operate as sign units for sonic generation and meaning is continually negotiated in the interplay among linguistic prompts, latent representations, and sonic realisation. Each output was thus approached as a contingent, situated articulation within a mutable representational field.

**2.2 Dialectical Assessment through Paired Études**

Central to the pedagogy was an étude-based assessment model comprising six bi-weekly assignments, each centred on a distinct AI modality. Every assignment paired a conventional étude—aimed at procedural compentence with the modality's intended operations—with an unconventional étude that deliberately misused or reframed the same system to expose its representational possibilities. The pairing cultivated operational fluency and deconstructive exploration in tandem.

---

[5] Cross-materialism refers to the process by which a singular source material—sonic, textual, or otherwise—is algorithmically processed and transformed across representational layers, resulting in novel sonic forms. The term describes how AI systems enable inputs such as vocal lines, environmental sounds, or instrumental gestures to be reconstituted into new auditory materials through latent manipulations. As used in this course, cross-materialism emphasizes the ontological fluidity and recombinant potential of materials traversed through latent space operations, offering an account of synthesis that emerges from material recomposition across representational domains.

[6] Semiotic slippage refers to moments in which the relationship between a signifier and its signified becomes unstable, producing shifts, ambiguities, or contradictions in meaning. In the context of sound and music, it describes instances where sonic signs evoke recognisable associations yet diverge from them through alterations in timbre, structure, or articulation, thereby opening space for new interpretations to emerge (cf. Barthes, 1977; Eco, 1976).

As an orienting example from the timbre transfer unit, the conventional étude asked students to produce sonically coherent transformations—applying instrumental or vocal characteristics to suitable sources to achieve convincing renderings, akin to the viral Ghostwriter track *Heart on My Sleeve*, which translated a producer's vocal into synthetic Drake/The Weeknd timbres.[7] By contrast, the unconventional étude treated timbre transfer as a *material disruptor* rather than a translator: students mapped orchestral articulations onto industrial noise, traced human utterances across metallic drones, or infused urban soundscapes with choral resonance. These cross-material recombinations activated latent space as an unstable, generative terrain, yielding hybrids that resisted neat stylistic taxonomy.

The text-to-audio unit further sharpened the dialectic. In the conventional étude, students crafted clear, contextually precise prompts to generate pieces whose structure and style aligned with their stated intentions (e.g., a chamber piece in the manner of Debussy or a trap beat in the style of Metro Boomin). Emphasis fell on controllability, structural coherence, and stylistic consistency with prompt engineering.

The unconventional étude re-positioned text-to-audio as a site of semantic and sonic volatility. Students wrote prompts that purposefully unsettled the relation between linguistic description and sonic realisation—introducing contradiction, polysemy, or self-negating constraints (e.g., "despaired joy, classical noise; nuclear test window shatter disco," or using ASCII emojis as prompts). Outputs typically occupied liminal aesthetic spaces: neither literal translations of the text nor detached from it, but suspended in semiotic slippage, where meaning remained in play and new forms could emerge. Iteration with identical prompts foregrounded non-repetition and stochastic drift, reinforcing an emergent, rather than templated, model of sonic flux.

By coupling normative and exploratory exercises, the assessment model enabled students to develop tool competence while interrogating the systems' embedded logics. Creative agency was trained as the capacity to reframe, reroute, and recompose expected behaviours—treating AI as a machinic interlocutor, and emerged through a willingness to test the elasticity of the system's interpretative capacity. In practice, this meant learning to work with intended affordances and to traverse failure modes deliberately, posing aesthetic questions, cultivating critical listening, and opening distinct trajectories for musical grammar.

**2.3 Independent, Student-Centered Learning and Course Assignments**

In contrast to didactic instruction or skill-oriented training, the course was deliberately structured around an open-ended, exploratory model that emphasized student autonomy and the cultivation of individual artistic trajectories. Assignments functioned as structured invitations for creative investigation rather than exercises with predetermined outcomes. Each assignment included a conventional application of the AI modality—encouraging technically proficient and outcome focused use of the tools—and an unconventional or deconstructive component that prompted students to challenge traditional approaches, exploring sonic ruptures, timbral anomalies, semantic drifts, and emergent expressive possibilities. This bifurcated approach created a reflexive space where students could simultaneously develop practical competencies while questioning fundamental assumptions about technological mediation and reimagining their relationship to musical creation through these tools.

---

[7] Ghostwriter's "Heart on My Sleeve" was a viral song released in April 2023 that used voice synthesis technology to translate the producer's vocal performances into synthetic voices closely mimicking Drake and The Weeknd. The track gained significant attention for its convincing vocal synthesis before being removed from streaming platforms due to copyright concerns, highlighting the complex legal and ethical questions surrounding AI voice translation that emulates specific artists. See: Sisario, B. (2023, April 18). An A.I. Hit With Fake Drake and The Weeknd Vocals Rattles the Music World. The New York Times. Retrieved from https://www.nytimes.com/2023/04/18/arts/music/ai-drake-the-weeknd-heart-on-my-sleeve.html

**Assignment sequence (six majors across the semester):**

1. **Symbolic composition.** AI-assisted MIDI/score generation and manipulation, treating MIDI notes as *matter* and their sound design/contextualisation as *form*.
2. **Timbre transfer.** Cross-material transformations between instrumental, vocal, and arbitrary sources, using timbre transfer to translate input audio into new sonic identities.
3. **Text-to-audio I.** Direct generation of audio from language prompts, treating words as sign units and prompts as assemblages of representational signs.
4. **Text-to-audio II ("rhizomixing").** Reshaping input audio through prompt engineering and TTA "remix" or "extend" functions to produce distinct or evolving forms.
5. **Text-to-audio III (audio for picture).** Generating TTA sound materials specifically for synchronisation and scoring of visual content.
6. **AI mixing & mastering.** Mixing and mastering an earlier work, or misusing a mixing/mastering tool to transform an existing sound recording.

Students concluded with a 20-minute class presentation. Those taking the course for 6 CP (rather than 3 CP for assignments + presentation) completed a practice-based research project consisting of a creative portfolio and a 1,000–4,000-word reflective essay that articulated the conceptual stakes, justified tool choices, and linked process to the theoretical frameworks discussed in class.

Importantly, each assignment was deliberately open in terms of aesthetic orientation and genre focus. Students could pursue sonic territories aligned with their interests while being encouraged to step beyond familiar idioms, engage critically with tool constraints, and "open their ears" to unfamiliar behaviours and latent materials. This balance of intentionality and exploration enabled a simultaneous practical and reflective stance toward every modality.

Pedagogically, the course functioned as a space for individuation: students developed distinctive modes of listening, conceptual framing, and aesthetic positioning. They learned to think—and hear—*with and through* the tools, attending to what AI systems afford, occlude, and reconfigure in relation to their evolving practice. By semester's end, most had expanded technical competencies while refining listening strategies and conceptual frameworks that integrated AI modalities into their work in distinctive ways.

This pedagogical structure proved especially fruitful in fostering diverse and unexpected outputs from students with widely varying backgrounds. Some participants entered with formal training in music theory, performance, or production; others arrived from disciplines such as architecture, media studies, and computer engineering. Notably, several of the most conceptually incisive and sonically distinctive projects came from students with limited prior music-making experience, whose approaches—less tethered to inherited stylistic templates—leveraged the assignments' exploratory brief to articulate compelling, idiosyncratic results.

**2.4 Aims and Learning Outcomes**

The course aimed to cultivate practical fluency and critical literacy across key AI modalities in music and sound—text-to-audio, neural audio synthesis, timbre transfer, symbolic composition, and related tools—within a coherent sequence of theoretical discussion, technical demonstration, case-study analysis, and guided experimentation. The goal was to enable hands-on competence while equipping students to interrogate the aesthetic, representational, and socio-cultural conditions embedded in these systems.

Creative work was framed as an inquiry-driven process. Students were consistently encouraged to engage in lateral thinking: borrowing concepts from sound studies, musicology, media theory, and philosophy to frame their creative decisions. Artistic case studies—ranging from Dadabots' neural death metal (2019), Oneohtrix Point Never's use of Riffusion and Adobe Enhance in Again (2023), and Patten's text-to-audio collage (2021) to Holly Herndon's "live training" performances (2019)—situated each modality within wider artistic practices and debates. moving beyond mere simulation of genre toward reconfiguring relations among technology, process, affect, and listening.

**By the end of the course, students were expected to:**
1. Demonstrate operational proficiency with core modalities (symbolic, timbral, vocal, text-to-audio, neural synthesis) in intended-use scenarios.
2. Exercise prompt and experiment design: formulate hypotheses, iterate systematically, and document process and results.
3. Practise critical listening and semiotic analysis, identifying phenomena such as timbral artefacts, semantic drift, and latent-space behaviours.
4. Exhibit medium awareness: analyse dataset priors, platform logics, and the mediating role of tools in shaping representation and reception.
5. Consider ethical and legal reflection (e.g., likeness, consent, provenance, attribution) into creative decision-making.
6. Communicate work effectively through presentations and reflective writing, linking creative outcomes to the course's theoretical frameworks.
7. Assemble a coherent creative portfolio that evidences both conventional and deconstructive engagements across modalities.

Crucially, the course did not aim to fix a single account of "how AI music works," but to foster reflexive, context-sensitive practice that can evolve as technologies change and their affordances shift. This demanded an attentiveness to the codified dimensions of music (notation, genre conventions, production workflows) *and* the open-ended, unstable territories these tools expose—where representation gives way to transformation, and emergent sonic forms arise from cross-modal entanglements. Through iterative practice and experimentation, students developed their own views and approaches to these technologies, building a foundation for continued exploration beyond the course's conclusion.

## 3. Course Structure and Content

Due to word count constraints, detailed information on the course structure, content, and assessment framework is provided in the Appendix, including a full outline of the course's sessions, assignment descriptions, and representative teaching materials.

## 4. Student Projects: Creative and Conceptual Explorations

A wide range of compelling works emerged from the course. The following case studies show how assignments operated as sites of material investigation, critical reflection, and a rethinking of tool–user relations. Each project probes the semantic, material, and affective tensions embedded in AI-based tools, often producing hybrids that cut across conventional categories of tool, genre, technique, and authorship.

**4.1 Excavating Latent Sonic Territories: Liam Fogarty's Étude for Timbre Transfer[8]**

Liam Fogarty reconceptualised IRCAM's RAVE[9] beyond timbral substitution, approaching latent space itself as a generative field for composition and material emergence. Working with the Vintage model, he devised a performance method that subverted standard transfer procedures: sustained tones from a Korg Minilogue XD were continuously modulated, then routed through RAVE to yield evolving gestures that hovered between recognisable categories (neither wholly synthesiser nor convincingly "vintage"). Rather than mimicking a target identity, the piece inhabits what Fogarty called the "amniotic fluid" of RAVE's latent space—a liminal zone of cross-material recombination.

```
Étude for Timbre Transfer (RAVE)
    1. VCO1 enters
        a. Play with cutoff and resonance
        b. Play with shape
    2. VCO2 enters
        a. Play with cross mod
    3. LFO enters (square wave mapped to cutoff)
        a. Play with LFO intensity & rate
    4. White Noise osc enters
        a. Play with noise shape
        b. Other voices fade out
        c. LFO fades out
```

**Figure 2.** Étude for Timbre Tranfer (RAVE) by Liam Fogarty.

Fogarty's work demonstrated an attentiveness to the non-linear behaviors and emergent properties of neural networks, privileging material exploration over representational fidelity. As he reflected in his accompanying essay:

"This assignment provided me with insight into the creative musical possibilities offered by timbral transfer models. Being initially surprised with the low fidelity of RAVE's outputs in regards to what the usual definition of timbral transfer suggests (a musically plausible translation from, for instance, my violin, to a 'vintage' melodic analogue), I instead turned to the aesthetics of the latent space itself." (Fogarty, 2024)

---

[8] The student's work can be accessed via the following digital repository:
https://drive.google.com/drive/folders/1xj2bhr6tQrhIOzMbjp9Ep4SIVb7hBu_f?usp=sharing

[9] RAVE (Realtime Audio Variational autoEncoder) is a neural audio synthesis framework developed at IRCAM, designed for high-quality, low-latency audio generation and manipulation in real time. Leveraging a variational autoencoder architecture optimized for efficiency, RAVE enables musicians and researchers to explore latent space transformations of sound for creative and performance contexts. Further information is available at: https://github.com/acids-ircam/RAVE.

Fogarty's étude exemplifies the course's insistence on engaging the material behaviours of AI tools: constraints are treated not as deficits but as productive conditions for discovery. By steering micro-modulations at the input to elicit macro-morphologies in the output, he effectively instrumentalised the model, using RAVE as a site for sounding latent structures rather than imitating a target source. Taken together, the work demonstrates latent-space materialism and the instrumentalisation of the model over timbral emulation.

**4.2 Sonic Palimpsests: David Bilek's Transformed Field Recordings[10]**

Inspired by acousmatic approaches by composers such as Bernard Parmegiani, and working from a single field recording made in Cairo, David Bilek's timbre transfer assignment processed the source recording through four RAVE models (ISIS, Percussion, MusicNet, Wheel), then aligned the parallel renderings to the original timeline. The result is a sonic palimpsest: layered temporal simultaneity and a stratified auditory in which original and transformed materials of the same event are folded together and coexist. Sometimes the transformed versions blend with the original; elsewhere they replace it, submerging the "real" soundscape beneath machine-mediated projections and producing an acoustic uncanny that estranges the familiar.

As Bilek articulates in his reflection:

"The composition reflects an exploration of reconstructing the eventfulness and liveliness of soundscapes through the use of different timbres. The Recombining of various models with the same source material—or blending one or more models with the original recording—alternate sonic realities are formed. These new versions, layered and mixed into the original, a weird feeling, where the real soundscape disappears behind the artificial ones. This process questions the authenticity and "realness" of the original recording, blurring the line between the real and the imagined." (Bilek, 2024)

Bilek's layering of parallel, model-derived renderings onto the original timeline produced a sort of palimpsest that suspends the recording's indexical claim and substitutes a machine-mediated topology**.** The documentary function is submerged beneath algorithmic projections; the familiar soundscape returns as an estranged double—an instance of what Fisher (2016) calls the *weird*:[11] the intrusion of that which does not belong and the collision of incommensurable frames. The ensuing perceptual destabilisation solicits an acousmatic stance**,** with the listener oscillating between attention to residual traces and immersion in spectral deformation, where authenticity is rendered negotiable rather than given. In this way, the étude demonstrates documentary destabilisation and an acousmatic literacy achieved through parallel timbre-transfer model renderings**.**

---

[10] The student's work can be accessed via the following digital repository:
https://drive.google.com/drive/folders/1lQvSxeXl6tHyfrygF9qy9LqvO68uAFuO?usp=sharing

[11] Following Mark Fisher (2016), the weird involves an encounter with something that fundamentally disrupts established ontological categories and expectations. It evokes an unsettling uncertainty about reality, marked by the presence of something that should not exist within normative structures. Central to the weird is the notion of an encounter with an entity or phenomenon that defies rational comprehension, thereby destabilizing the boundaries between the known and the unknowable. Drawing on the works of H.P. Lovecraft and speculative fiction, Fisher argues that the weird signals an intrusion of an "outside," destabilizing the subject's sense of reality through alterity. Fisher's concepts of the weird and the eerie were discussed in class as part of our theoretical framing.

## 4.3 Counterfactual Sampling: Laura Bouget's presentation *Who Owns Sound? AI, Sampling, and the Question of Authorship*[12]

Laura Bouget's presentation, *Who Owns Sound? AI, Sampling, and the Question of Authorship*, offered a critically astute and culturally situated investigation into authorship, appropriation, and identity in AI-mediated sampling practices. Her presentation approached Udio (a text-to-audio model) as a medium of critical intervention—a way to reflect on and recompose the materials that it implicitly encodes.

Central to her presentation was a striking gesture: she took Ye's (Kanye West) *My Song* (a track built around a sample from Labi Siffre's original composition), and used Udio's prompt and lyric interface to generate a fictive version in which Siffre himself sings back against the appropriation. The resulting track stages a ghosted reclamation, with the generated voice lamenting, "This is not my song / Kanye took it away / I should go to court," effectively reversing the directional flow of authorship through a simulacrum of the sampled artist reclaiming agency within the derivative work.

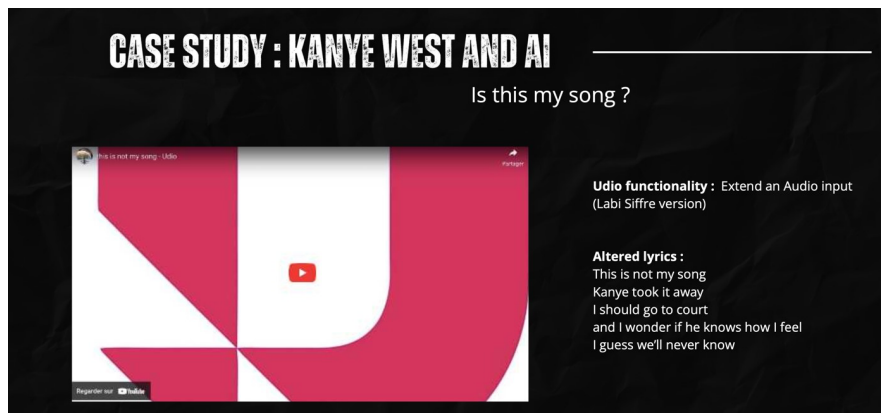

**Figure 3.** Bouget's AI-generated counterfactual vocal response from the perspective of the sampled artist.

Conceptually, the presentation interrogated the intersections of sampling culture, stylistic inference, and AI "spawning," foregrounding questions of sonic ownership and cultural legacy in the age of generative media. This work engaged the platform's interplay of sampling, spawning, and stylistic inference while offering a pointed commentary on the broader implications of sampling culture in the age of generative AI—addressin the cultural politics of sonic ownership and creative attribution in contemporary media landscapes, using AI to question and invert historical dynamics of appropriation. Bouget's counterfactual studies mobilise text-to-audio for performative critique: authorship is inverted, lineage speaks back, and platform affordances become dramaturgical devices that can sing in known voices. By exploiting TTA's singing/identity capacities (lyric prompts, vocal overlays, remix/extend), the work exposes how platform architectures can mediate attribution, provenance, and cultural memory. In this way, the project demonstrates authorship inversion, the critical use of TTA's vocal identity affordances, and a platform-aware critique within TTA workflows.

---

[12] The student's work can be accessed via the following digital repository:
https://drive.google.com/drive/folders/1PvWPhA0EcwCvZ8PsgvIi8tmdKwnZOZhq?usp=sharing

**4.4 Khalif Shirad's Presentation** *Musical Decomposition: Rhizomatic Deterritorialization in AI Music Composition*[13]

Khalif Shirad's presentation, titled *Musical Decomposition: Rhizomatic Deterritorialization in AI Music Composition*, offers a conceptually ambitious inquiry into the aesthetic limits of prompt-based generation. Although his listening practices are rooted in pop and EDM (with Lady Gaga as a key reference), he intentionally traversed different terrains, drawing on Deleuzian deterritorialisation/reterritorialisation, artworks such as The Caretaker, and Laura Bouget's TTA sampling studies to frame AI as a method of deliberate unmaking rather than stylistic reproduction.

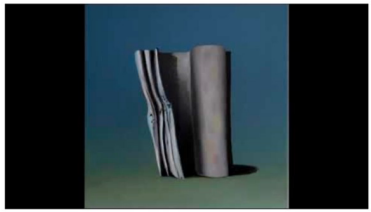

**Figure 4.** Conceptual frameworks and artistic precedents informing Khalif Shirad's analytical approach to AI-mediated sonic decomposition.

The presentation centred on the progressive defamiliarization and decomposition of source material through Udio's remix and extension functions, guided by ChatGPT-generated prompts. Beginning with a recognizable track (Katy Perry's *California Gurls*), Shirad applied a sequence of rhizomatic extensions that displaced the original's affective coordinates. Layers of noise, abrasive textures, ambient deconstructions, and abrupt genre collisions eroded the sonic identity of the source, with each recombination catalyzing further drift, and pushing the material away from recognizability toward abstraction.

---

[13] The student's work can be accessed via the following digital repository:
https://drive.google.com/drive/folders/1xL28eeWf-IFSUqFjE0OR17wOlYEFDPC0?usp=drive_link

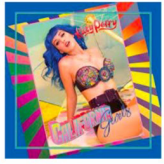
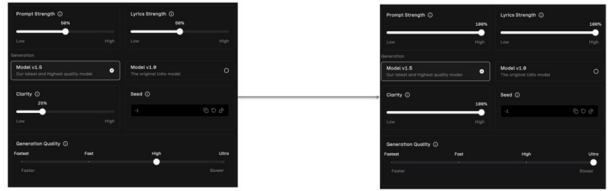
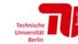

**Figure 5.** Initial methodological framework employed in Shirad's deconstruction of normative audio representations through text-to-audio transformation.

His methodology explicitly targeted moments of representational breakdown, perceptual disjunction, and algorithmic indeterminacy rather than seeking fidelity or coherence. The project's strength lies in its sonic results alongside its rigorous conceptual framing and effective prompt-based methodology for arriving at its desired aesthetic outcome. Khalif explicitly theorized the notion of flawed representation in AI as an aesthetic condition ripe for creative exploitation, demonstrating how these deviations could be systematically explored as a form of compositional practice.

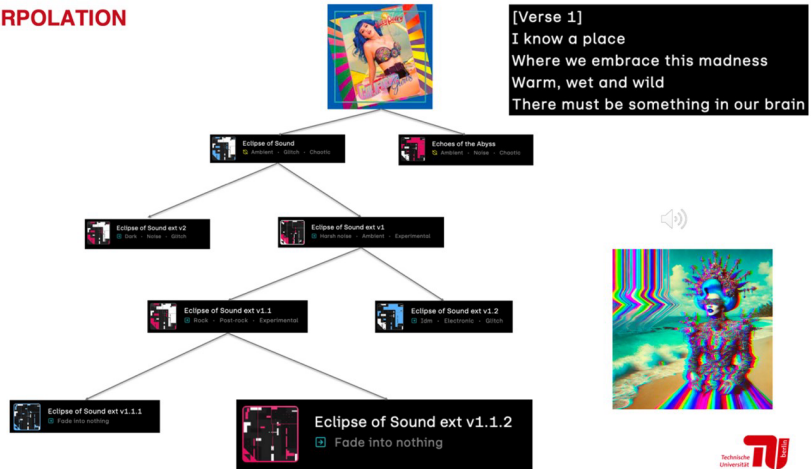
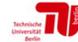

**Figure 6.** Second phase of Shirad's systematic approach to rhizomatic deterritorialization utilizing combinatorial prompt engineering.

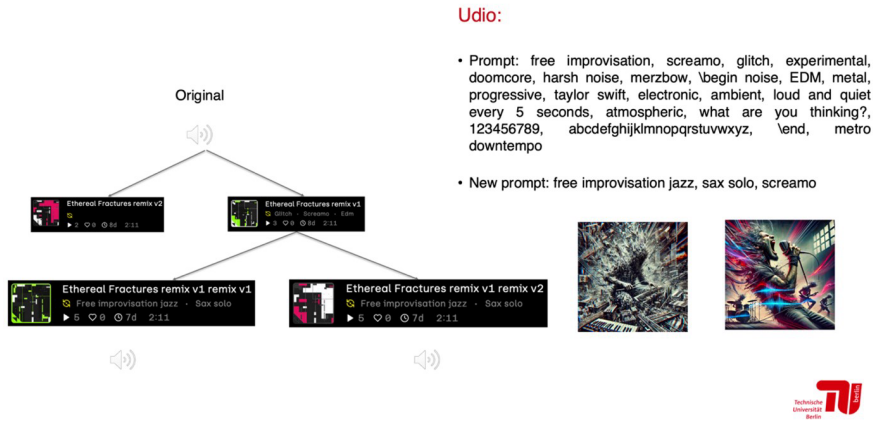

**Figure 7.** Final methodological stage demonstrating recursive application of semantic drift to achieve progressive abstraction of source material.

This study shows that text-to-audio systems such as Udio can serve as instruments for systematic aesthetic deconstruction when handled with conceptual rigour. Through recursive prompt design and the platform's remix/extend affordances, Shirad drives progressive material dissolution**,** demonstrating capacities for experimental composition beyond genre replication. The work highlights the capacity of contemporary AI music systems to support complex creative investigations that bridge personal musical sensibilities with "avant-garde" compositional strategies, opening new possibilities for how generative technologies might be integrated into experimental artistic practices. The project also couples a pop-oriented sensibility with avant-garde logics of erosion and abstraction, yielding reproducible procedures for research-creation. In sum, it delineates durable methods for integrating contemporary TTA workflows into deconstructive artistic practice.

**4.5 Intersemiotic Translation: Felix Luis's Cetacean-Human Musical Dialogues[14]**

Felix Luis's presentation pursued a cross-material exploration between human and non-human expression via AI-mediated transformation. Using whale songs as input audio, he examined how Udio can mediate across divergent sonic ontologies—the bioacoustic communication of marine mammals and the culturally codified structures of human musical traditions. The choice of cetacean material was conceptually pointed: its inherent melodic/rhythmic organisation offered a fertile substrate for testing deterritorialisation/reterritorialisation and for probing how a TTA model recontextualises source signals.

The project presented a triptych of études, each applying the same whale recording to a different prompt: "Avant-Folk, acoustic, improvisation"; "Gustav Mahler, symphony, Romanticism"; and the deliberately open "Human Music." The first étude rendered the vocalisations into an experimental folk idiom, preserving organic gestural qualities while situating them within improvisatory practice (with prompt *metatags* from Hermeto Pascoal RateYourMusic page). The second enacted a more forceful reterritorialisation, aligning biological utterance with the timbral/formal logics of nineteenth-century orchestral writing. The third, referencing *Rick and Morty*, served as a humorous diagnostic of model priors: when prompted with an open category like "Human Music," the system exposed its own latent assumptions about what constitutes the musical "universal."

---

[14] The student's work can be accessed via the following digital repository:
https://drive.google.com/drive/folders/1tzy87XV7yTPaKCN-FFf9OxwdLvqEsSr0?usp=drive_link

The third étude, prompted by the intentionally open-ended term "Human Music" (a reference to the *Rick and Morty* television series), provided insight into Udio's latent conception of musical universal—particularly noteworthy given that all training data in Udio basically represents "human music". The resulting output offered a compelling glimpse into how the model synthesizes and prioritizes musical elements when given such a prompt, and this experiment illuminated how particular prompts can surface the model's underlying categorical organization of musical attributes and aesthetic hierarchies.

Through these three studies, Luis treated text-to-audio as an intersemiotic processor and transmodal conduit**:** prompts condition how signs migrate between regimes, and varying degrees of semantic specificity yield correspondingly different stabilisations of form. Taken together, the work demonstrates intersemiotic translation, prompt-conditioned reterritorialisation, and the surfacing of model priors when semantic constraints are relaxed**.** It also shows Udio's strong capacity to transform input materials according to representational prompts, achieving timbre-analogous mappings that read musically—*e.g., a whale's gushing articulated as an orchestral ride/splash gesture*.

### 4.6 Associative Spawning and Beat-Tape composition: Ben Klaster's Final Creative project *Invisible Signs*[15]

Ben Klaster's final project, *Invisible Signs*, assembles a nine-part beat tape entirely from Udio generations, using Italo Calvino's *Invisible Cities* as the sole prompt source and structural armature. He frames text-to-audio as an *intersemiotic, transmodal conduit*—a system that translates meaning across sign systems and enables rapid associative processes between words and sounds. The project treats these associative embeddings as creative affordances while acknowledging that a model's associations will diverge from any single listener's own.

To minimise lyrical leakage and keep the system focused on sonic interpretation, Klaster worked in Udio's Manual Mode with the Instrumental option, generating *two outputs per prompt*. Each prompt combined the city's name with a thematic heading and a cluster of phrases drawn from (or inspired by) the text—sourced from a German translation—thus encoding his interpretive stance directly into the semantic scaffold. He then assembled nine pieces—one per chapter—in Bitwig Studio: importing all generations; selecting at least one sound object per prompt; setting tempo from rhythmically clear anchors; time-stretching with Elastique Eco; occasional pitch shifts for tonal compatibility; light EQ to avoid clashes; and a simple mastering chain (compressor, multiband rack, stereo-width tool, limiter) to raise level. Transitions are shaped with tempo automation, preserving a beat-tape flow.

The work stages a multilayer translation chain: Calvino's prose → Klaster's interpretive prompt → model's latent audio realisation → sample-based recombination in a plunderphonic manner. Klaster explicitly situates the tape at the interface of *sampling* and *spawning*: the source material is spawned (AI-generated in the image of the training corpus), but the composition process is classic collage—cutting, layering, and recontextualising short fragments into affectively coherent tracks. He contrasts this with Plunderphonics' reliance on recognisable excerpts, noting that spawned materials produce stylistic spectres rather than quotable recordings; the listener readily identifies genres (folk, metal, hip-hop, jazz) but cannot recognise a specific sample, since none previously existed.

---

[15]The student's work can be accessed via the following digital repository:
 https://drive.google.com/file/d/1WGVdu8h43GAioiumMOt7LOoK1VtspW9l/view?usp=sharing

Because prompts intentionally avoided straightforward musical descriptors, Udio's responses often revealed the model's own associative priors (e.g., "memory" tending toward nostalgic timbres; "signs" yielding digital idioms), though Klaster emphasises that consistent patterns were difficult to establish without a larger prompt study. The tape's cohesion emerges ex post through editorial constraints—tempo, spectral complementarity, and transitional kinships—rather than from any one genre.

*Invisible Signs* demonstrates (i) associative spawning driven by literary prompts; (ii) prompt-conditioned selection and beat-tape assembly as a reproducible workflow; and (iii) a clear analytic distinction between sampling practices and spawning-derived source pools, where genre legibility persists without quotation. In short, Klaster shows how TTA outputs can be composed like samples while functioning as spectres, yielding a cohesive beat tape from entirely AI-spawned material.

## 5. Reflections: From Skepticism to Creative Co-articulation

Many students entered the course with measured scepticism—some driven by ethical concerns, others doubtful about the aesthetic or technical merit of AI-generated material. These stances shifted early in the semester as theas the course introduced carefully selected case studies and demonstrations that challenged reductive notions of automation or creative replacement, and framed AI as a contingent, coded, and interpretable medium— capable of revealing latent structures and inviting speculative forms of musicking.. Through hands-on engagement with tool and post-structuralist framing, students moved from caution to curiosity and, in many cases, to a genuine sense of curiosity and creative investment.

### 5.1 AI as Epistemic Infrastructure and Meta-awareness Device

Across modalities—voice synthesis, timbre transfer, neural synthesis, and text-to-audio—students came to understand AI less as a set of replication tools and more as a mediating environment through which ideas are translated, reframed, and reimagined. In this role, AI functioned as a meta-awareness device: it sonified genre priors, timbral defaults, loudness norms, and musical codifications, thereby rendering audible cultural assumptions that typically pass as neutral. This reflexive capacity echoes Bourdieu's account of symbolic violence, where dominant cultural classifications are misrecognised as neutral or universal rather than as products of specific power relations (Bourdieu, 1991). Treating AI tools as epistemic instruments expanded students' critical literacy (not only about AI but about music's cultural construction), positioning them to question how infrastructures shape listening, composition, and authorship.

### 5.2 Interpretative Multiplicity and Process-based Inquiry

The paired-études structure cultivated a disposition to treat ambiguity as resource. Students learned to leverage interpretative multiplicity—conflicting outputs, unstable behaviours, and generative "imperfections"—as stimuli for analysis and invention rather than as errors to correct. Latent space became a creative terrain, and composition assumed an investigative character attentive to semantic slippage and emergent form. This embrace of epistemic indeterminacy resonates (and was partially inspired) with Rancière's figure of the ignorant schoolmaster: learning through experimentation and discovery rather than the transmission of predetermined knowledge (Rancière, 2010).

### 5.3 Recursive Tool Engagement and Creative Co-construction with LLMs

A notable pattern was students' widespread integration of LLMs into creative workflows, using them to explore prompt phrasing, conceptual framing, and philosophical positioning—prompting the prompts that would later inform their generative outputs. This recursive engagement with tools reflects a broader shift in contemporary creative practice, where meaning is increasingly co-constructed through algorithmic mediation. Musical creativity in AI-mediated contexts increasingly manifests through thinking with and through machines—developing ideas, methods, and sensibilities are developed *with and through* computational affordances and their interplays.

**5.4 Implications for Critical AI Pedagogy**

This pedagogy reveals AI's still-underused value for critical artistic education beyond established codifications and commercial templates. Rather than training students to become efficient operators of pre-packaged tools, the course positions AI systems as sites of cultural investigation and aesthetic experimentation**.** Treating models as media—historically situated, corporately maintained, and saturated with priors—makes them examinable rather than merely usable. In this frame, creative work proceeds as inquiry: students learn to surface the biases, affordances, limits, and unexpected capacities of the systems through which they compose.

A critical AI pedagogy should also contest the subtle "indoctrinations" of music education: inherited assumptions about what counts as music, how it should sound, and what tools ought to do (e.g., reproduce genre norms, increase efficiency, deliver fidelity). New technologies have always introduced distinct aesthetics and sonic forms**;** AI is no exception. Yet contemporary discourse often ignores these emergent materialities, either celebrating frictionless productivity or fearing collapse into pastiche. I argue for cultivating dispositions that hear and think with différance—that is, attuned to deferred, relational meaning and productive misalignment—recognising AI's distinctive affordances (intersemiotic translation across text/symbol/timbre, latent-space recombination, stochastic variation, prompt-conditioned form) and engaging them as compositional resources rather than as threats to a pre-given musical order.

At the same time, the risk of homogenisation is real: model priors and platform incentives can pull outputs toward central tendencies; recommender ecologies reward predictability; "best-practice" prompt recipes codify style. A critical pedagogy must therefore teach techniques of divergence alongside fluency, so students can hear and work against these convergences.

Concretely, this implies a set of pedagogical commitments:

- **Reveal and interrogate structural priors (before datasets).** Begin from structuralism: treat music as a codified sign-system organised along paradigmatic choices and syntagmatic combinations—tonal and metric schemata, genre grammars, orchestration and studio idioms, DAW-era workflow routines, and habituated listening that conditions expectation. Surface how training, repertoire, and institutional assessment naturalise these codes—the *indoctrination of music*—and how AI both reproduces and perturbs them. Have students map which codes are invoked, stabilised, or subverted in each experiment, and design practices that work *with,* *bend,* or *break* such structures.
- **Design for divergence.** Pair intended-use études with deconstructive études that traverse failure modes (misuse, contradiction, semantic drift), explicitly seeking non-convergent results.
- **Foreground intersemiotic practice.** Assign cross-modal tasks (text→audio, audio→textual description, symbol→timbre) to cultivate sensitivity to translation, loss, and invention.
- **Document the process.** Make prompt design, parameter settings, iterations, and listening notes examinable artefacts; assess method and reflection, not only outputs.
- **Teach critical listening.** Train students to detect genre priors, loudness norms, timbral defaults, artefacts, and "model signatures," and to articulate their implications.
- **Integrate ethics and law as design constraints.** Likeness, consent, provenance, and attribution are not afterthoughts but compositional parameters that shape what should be made.

- **Interrogate platforms.** Analyse business models, moderation regimes, and metric incentives; relate platform logics and their dispositif to aesthetic outcomes and labour relations.
- **Encourage repertoire-building with the new.** Curate case studies that highlight AI's distinctive aesthetics (e.g., latent-space materialism, counterfactual sampling, prompt-conditioned continuation) so students hear possibilities they did not know to seek.
- **Assess for exploratory competence.** Reward strategies that purposefully stretch or repurpose tools, not just polished genre compliance.

Such approaches are particularly urgent as AI systems become institutionalised across educational, professional, and creative domains. Without critical frameworks that treat these technologies as cultural artefacts and infrastructures, adoption risks reproducing existing power structures and aesthetic hierarchies—exactly the homogenising tendency many fear. By coupling post-structuralist theory with medium awareness and deconstructive practice, the pedagogy advanced here offers one model for integrating AI into creative education in a way that expands rather than erodes critical thinking, cultural literacy, and the capacity to steward emerging sonic forms.

## Conclusion

This paper presented a practice-based, critically reflexive pedagogy for *AI in Music and Sound: Modalities, Tools, and Creative Applications*. Against a landscape in which academic approaches to AI in music often remain either overly technical orexclusively theoretical according to established paradigms, this course proposed a curriculum that treats AI as a transmodal conduit and a mediating environment**,** coupling operational proficiency with conceptual interrogation. Grounded in medium theory, structuralist and post-structuralist inquiry, and deconstructive practice, the design positioned AI systems as sites where meaning is translated, perturbed, and reconfigured—rather than as neutral utilities or mere production aids that follow established conventions of musical production and normative aesthetic judgments..

The *paired-études* assessment model proved central. By yoking intended-use études to deliberately reframed or "misused" études, students learned to work *with* affordances and *against* them, cultivating method, critical listening, and process-aware composition. Case studies demonstrated recurrent patterns—latent-space materialism, documentary destabilisation, authorship inversion, prompt-conditioned continuation—showing how intersemiotic translation (text↔symbol↔timbre↔audio) can be mobilised as a compositional resource. In parallel, students developed medium awareness**:** an ability to hear and analyse how structural priors (tonal/metric schemata, genre grammars, orchestration and studio idioms, DAW workflows) and platform priors (datasets, architectures, interface constraints) shape what becomes audible.

Pedagogically, the approach moved students from initial scepticism to creative co-articulation with computational systems: thinking and hearing *with différance*—attuned to deferred, relational meaning and productive misalignment—rather than seeking frictionless replication. The resultant learning dispositions favour inquiry over recipe, method over mimicry, and reflexive documentation over opaque "best practice."

The implications for critical AI pedagogy are twofold. First, higher education can and should contest the subtle indoctrinations of music—assumptions about what counts as music, how it ought to sound, and what tools should do—by revealing, testing, and sometimes breaking structural codes. Second, AI introduces distinctive aesthetic capacities (e.g., latent recombination, stochastic variation, cross-modal prompting) that warrant curricular attention in their own right. Without such frameworks, institutional adoption is likely to intensify homogenisation pressures embedded in platform logics and recommendation ecologies. With them, AI becomes an object of study and a partner in experiment—an infrastructure for cultural analysis as well as creative invention.

In sum, the framework offered here integrates conceptual thought, artistic practice, and technological inquiry to prepare students for a world increasingly mediated by AI. It models how courses can move beyond a binary of acceptance or rejection and instead cultivate durable methods for listening to, working through, and arguing with computational media—methods that expand critical literacy while opening genuinely distinct (and productively *different*) trajectories for musical creation and reflection. Crucially, it also equips students for active participation in shaping how AI is understood, developed, and deployed within creative communities—as critical users, informed interlocutors for designers, and responsible co-authors of emerging practices.

**Ethics Statement**
This research was conducted in accordance with the ethical guidelines of the institution where the course was taught. All student work featured in the case studies is included with explicit permission from the students, who were informed about the research aims and publication intentions. Students retained full copyright and creative ownership of their works, and their contributions are fully acknowledged throughout. No personal data beyond academic outputs and reflections were collected or analyzed. No conflicts of interest exist in the presentation of this pedagogical research.

This paper was written with the assistance of large language models (LLMs), which were used to help refine prose, improve clarity, and enhance the articulation of pedagogical concepts and theoretical frameworks. The ideas, arguments, and pedagogical insights presented in this paper are entirely those of the author; no content was generated by LLMs independently.

# Appendix

Included in the appendix folder are selected materials from the course *AI in Music and Sound: Modalities, Tools, and Creative Applications*. These comprise official session slides from various weeks, full assignment sheets (including guidelines and tool suggestions), the course assessment components document outlining evaluation criteria, and the university's official course description. These materials collectively offer a view into the course's pedagogical architecture and instructional intent. They are accessible via the following link: https://drive.google.com/drive/folders/1MDZIHEN6dXUj1WMve9ASbC8xrkr0nj8k?usp=sharing

## Appendix A. Course Structure and Content

### A.1 Symbolic Composition and MIDI-Based Systems

The initial sessions introduced students to AI-based composition through symbolic and MIDI-based systems. These early explorations foregrounded questions of pattern recognition, stylistic modeling, and the operational logics of training data. Drawing from structuralist and post-structuralist perspectives, the course invited students to interrogate how such systems instantiate codified notions of musical form and to consider how alternative compositional strategies might emerge from engagement with their internal biases and limitations. The sessions acknowledged music as a "medium", and as an established codified representational language with its own syntactic and semantic structures while examining how AI systems both reproduce and potentially transform these conventions.

From the outset, the course established its signature pedagogical approach through paired conventional and unconventional études, designed to cultivate technical competency and critical investigation of AI systems' embedded assumptions. For instance, in the symbolic composition assignment, the conventional étude tasked students with using MIDI-based AI models (such as Google's Bach Doodle) to harmonize melodies or generate counterpoint in the recognizable style of J.S. Bach, demonstrating technical proficiency with baroque compositional conventions and voice leading principles. The unconventional étude challenged students to repurpose these same Bach-trained models for entirely different creative purposes—using the harmonic progressions as rhythmic triggers, extracting melodic fragments for sample triggering through MIDI, or applying the baroque voice leading to generate material for contemporary genres such as trance music or plunderphonics compositions. This approach revealed how AI models trained on specific styles could be creatively misappropriated to produce aesthetically divergent outcomes.

Additionally, the philosophical concept of hylomorphism was introduced as a framework for understanding the relationship between MIDI data (notes as matter) and sonic realization (sound design and instruments as form), demonstrating how a single melodic sequence could manifest in radically different tempi, timbres, and textures, thus revealing multiple possibilities for musical worldmaking. These discussions positioned music as a medium with its own materiality and considered the implications of AI's translation of symbolic abstractions into sonic material. This dual-étude methodology served as a practical introduction to fundamental concepts of representation and semiotics, enabling students to observe firsthand how the same symbolic content could be decoded and recontextualized across different aesthetic frameworks, thereby laying essential groundwork for the more advanced theoretical discussions that would follow throughout the course.

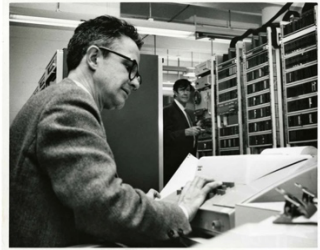
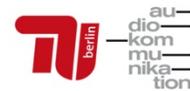

**Figure A1.1.** Slide framing music as a codified sign-system to motivate the symbolic/MIDI strand.

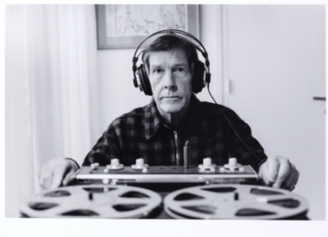
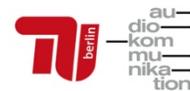

**Figure A1.2.** Slide framing music as an aesthetic experience.

**A.2 Live Performance and Embodied Interaction**

Subsequent sessions examined the domains of live performance and embodied interaction. The course considered improvisation systems and machine learning instruments, situating these tools within broader discourses on performativity, liveness, improvisation, and the reconfiguration of musical agency. In this context, AI was framed as an actant, with students encouraged to conceptualize interaction, unpredictability, and material feedback as central compositional elements. Case studies from George E. Lewis and Franziska Schroeder (and others) informed these discussions, particularly regarding the heuristic circle of improvisation and Actor-Network Theory as applied to human-machine interaction.

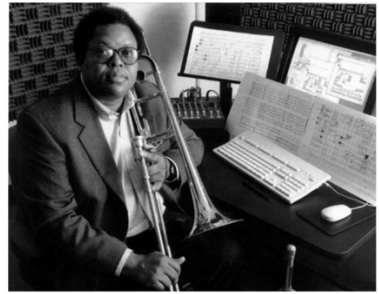

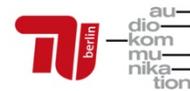

**Figure A.2.1.** Slide framing machine subjectivities in improvisation. AI as an actant that stakes musical territory, responds to conditions, and asserts positions within human–machine ecologies (after George E. Lewis).

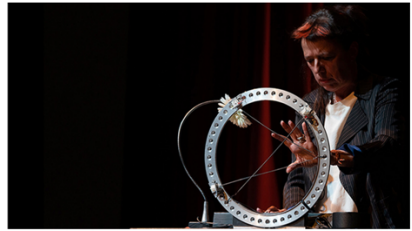

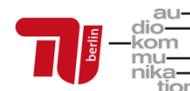

**Slide A.2.2.** Slide framing performance as discovery with machine learning—embracing adaptive, exploratory behaviour over established outcomes, where instruments surprise rather than merely execute (after Laetitia Sonami).

**A.3 Voice Synthesis and Timbre Transfer**

The following sessions focused on voice synthesis and timbre transfer, utilizing autoencoder architectures and latent space as conceptual scaffolding. Students examined how voice models simulate, recombine, and abstract the human voice as forms of sonic identity and expressive material. A two-part assignment structure was again implemented: first, students produced a conventional vocal étude by generating a synthetic singing voice from textual prompts or input audio; subsequently, in an unconventional variation, they were challenged to treat the vocal model as a generator of alternative sonic forms—reprocessing vocal timbres into rhythmic textures, drones, environmental artifacts and so on. Building upon the concept that voice synthesis could generate diverse sonic forms beyond vocal performances, in a second assignment students further explored timbre transfer through various instrumental representations using tools such as Neutone Morpho, IRCAM's RAVE, and Magenta's DDSP. This bifurcated pedagogical structure exemplified the course's design philosophy: exposing students to the operational affordances of each modality while simultaneously encouraging experimental deviation and conceptual reconfiguration.

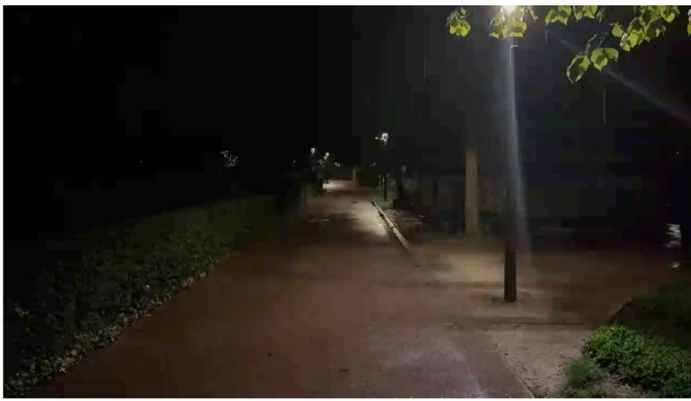

**Figure A.3.1.** Slide framing simulation, simulacra, and hyperreality (after Baudrillard) through AI voice models—synthetic vocal likenesses that simulate, substitute for, and sometimes eclipse the referent, problematising indexicality and authorship.

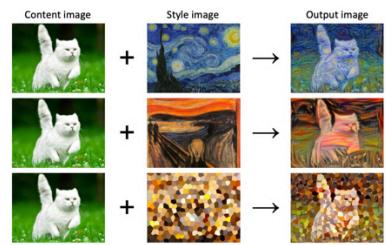

**Figure A.3.2.** Slide framing timbre transfer as an interstitial/intermediary practice—translation

across representational regimes in which outputs occupy a liminal zone between source materiality and target identity.

**A.4 Text-to-Audio as Primary Modality**

The course subsequently turned to neural audio synthesis and text-to-audio paradigms, with text-to-audio (TTA) emerging as the primary focus and most extensively explored modality of the course. Students engaged with systems capable of translating textual descriptors into complete sonic outputs, exploring the rhetorical, connotative, and semiotic implications of prompt design. Throughout these sessions, the course elucidated the complex intersemiotic processes underlying how these systems generate audio materials, developing a critical vocabulary to approach these technologies. The course conceptualized words as catalytic sign units and prompts as assemblages that trigger latent sonic explorations. Students reflected on the tension between control, association and indeterminacy, the rhetorical performativity of textual inputs, and the ways meaning is constructed through sonic material that resists straightforward representation.

Students first crafted prompt-driven études using stylistic or genre-based descriptors through both conventional semantic combinations (like "dark ambient electronic") and unconventional juxtapositions (such as "overwhelming angry joy"), exploring how different textual associations navigate and shape the resulting sonic territories. A crucial component involved learning how to reach particular cartographic areas of these models' vast latent spaces and generate material within specific aesthetic territories through strategic prompt engineering.

In subsequent assignments, students approached text-to-audio tools through remix logics—inputting audio material alongside conceptual prompts to explore how prior sonic identity could be transformed, displaced, or re-signified. In platforms like Udio, any generated sound object can be endlessly remixed and reprocessed, creating recursive cycles of material transformation that function as springboards for further rhizomatic exploration. This capacity for endless recombination revealed how each sonic output becomes simultaneously an endpoint and a starting point for possibel creative trajectories. Deleuze and Guattari's theoretical concepts of assemblage, rhizome, and reterritorialization provided crucial frameworks for understanding these processes of remixing, fragmentation, and non-linear compositional strategies with TTA. Students were encouraged to conceptualize their works as nodes within larger conceptual and material networks, shaped by ongoing processes of recombination and recontextualization, rather than as singular artifacts with fixed identities.

Across these sessions we emphasised the tools' expressive capacities alongisde their dispositif—the assemblage of cultural expectations, technical constraints, corporate incentives, and user behaviours that shape how they are imagined and used. Through sustained engagement, students moved from tentative operation to articulated positions on when and how such systems should be employed, with what limits, and to what aesthetic or ethical ends. They developed theoretical orientations and working compositional grammars, treating text-to-audio as a site of exploration, productive misalignment, emergence, and aesthetic inference rather than a reductive shortcut to instant sound objects.

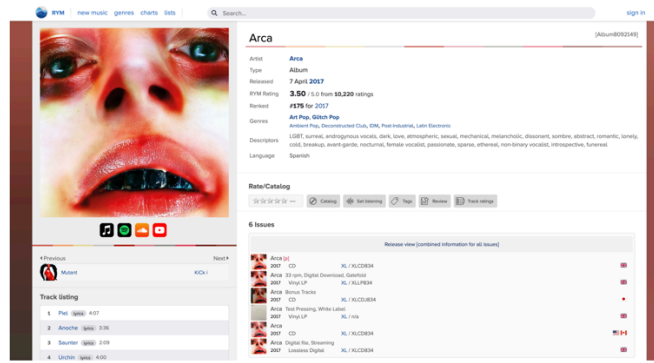
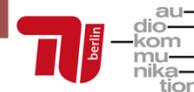

**Figure A.4.1.** Slide demonstrating the use of genre/mood tags and metatag vocabularies (from platform metadata) as semantic scaffolds to steer TTA prompts toward targeted regions of latent space.

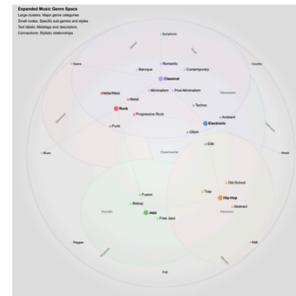
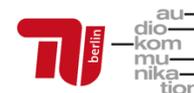

**Figure A.4.2.** Slide showing words as rhizomatic sign-units—single tokens and their combinations act as semantic triggers/catalysts that produce associations and dissociations in latent space, yielding emergent prompt behaviours.

**A.5 Text-to-Audio for Visual Content**

In the final weeks, the course explored two culminating applications of AI in sonic creation. Students used text-to-audio tools such as Udio and ElevenLabs to score and sound design visual content, examining the interpretative interplays between textual prompts and resulting soundscapes through semiotic interaction—translating visual narrative elements into textual prompts that generate appropriate sonic accompaniment. The class explored various audio-to-visual aesthetic approaches, from atmospheric underscoring to synchronized sound design, investigating how AI-mediated audio could transform, complement, or recontextualize visual narratives.

Maintaining the course's dual-étude structure, students approached this assignment through both conventional and unconventional methodologies. In conventional approaches, students crafted text prompts that directly mirrored the visual content's descriptive elements or emotional qualities—generating ambient soundscapes for nature footage or rhythmic compositions for urban scenes. The unconventional études challenged students to employ divergent or deliberately mismatched prompts to score the same visual material, exploring how semantic disjunction between audio and visual elements could create distinct interpretative possibilities and challenge conventional audiovisual relationships.

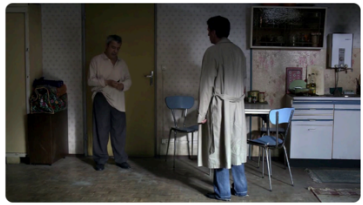

**Figure A.5.1.** Slide showing AI-assisted scene analysis (ChatGPT) to derive cue lists from a visual still, informing text-to-audio scoring prompts.

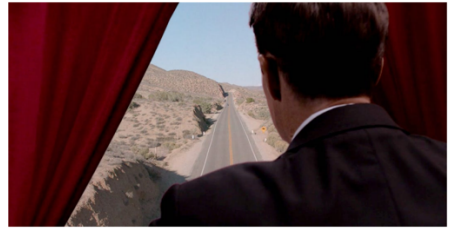

- **Associative approaches:** Matching specific sonic textures with visual elements.

- **Juxtapositional approaches:** Using contrasting sounds to create tension or subvert expectations.

This process aligns with Roland Barthes' idea of the *readerly* and *writerly* text (S/Z, 1970): the output can be experienced as either **closed** (fixed in its meaning) or **open** (inviting interpretation and interaction).

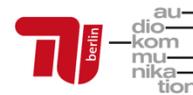

**Figure A.5.2.** Slide contrasting associative (alignment) and juxtapositional (contrast) scoring strategies, linked to Barthes's readerly/writerly distinction.

**A.6 AI Mixing and Mastering**

Subsequently, the course addressed AI-driven mixing and mastering, reflecting on how these systems interface with the standardization of sonic aesthetics in contemporary production cultures. The concluding assignment invited students to revisit a previous work and apply AI mixing tools to shape its final form. Critically, students were asked to reflect on the technical aspects of their decisions alongside how these systems encode assumptions about genre, loudness, clarity, and taste—revealing how AI participates in the constitution of sonic norms.

The conventional étude focused on addressing particular audio issues or achieving polished, industry-standard results using AI tools for their intended purposes—applying EQ, compression, and spatial effects to enhance clarity and commercial viability. The unconventional étude encouraged creative misappropriation of these same tools, such as using AI denoising algorithms as creative processors rather than corrective devices, exploiting their artifacts and limitations to generate novel timbral textures and spatial distortions that transformed the source material in unexpected ways.

## Artistic Case Study: OPN
Adobe Enhance on Instruments: Oneohtrix Point Never – Again (2023)

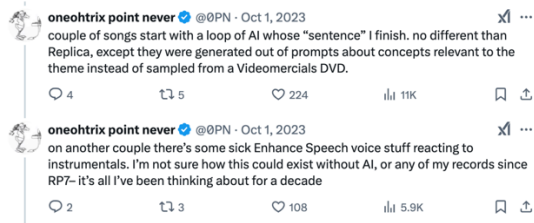

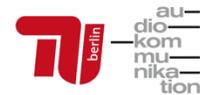

**Figure A.6.1.** Slide documenting OPN's use of AI in the album *Again* (2023).

## Artistic Case Study: Amnesia Scanner and Freeka Test
Deconstructive Sound Processing: FT Disco (Active Noise Cancelling Script) (2024)

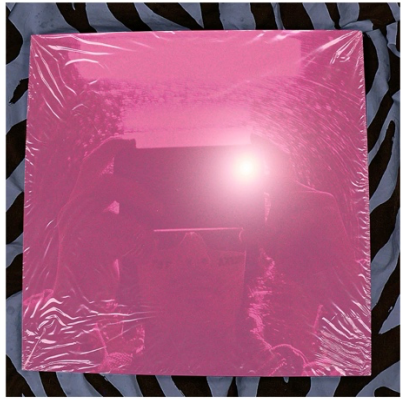

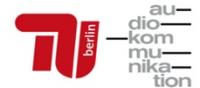

**Figure A.6.2.** Slide illustrating deconstructive processing via an active noise-cancelling script that algorithmically subtracts/negates signal components in the album *Hoax* (2024).

Collectively, the course modules constituted a comprehensive investigation into AI modalities in music and sound creation. Each session integrated theoretical foundations, philosophical concepts, artistic case studies, in-class demonstrations and guided experimentation, supporting a pedagogy that positioned AI systems as objects of critical inquiry and creative catalysis.

**A.7 Speculative Futures: The Age of Hyperreproduction**

The course concluded with a speculative discussion on the emergent "age of hyperreproduction," a cultural and technological condition in which generative systems exponentially multiply and mutate existing materials.

Students considered how AI's capacity for rapid, large-scale reproduction destabilizes notions of originality, authorship, and labor, creating an environment where iteration and endless content production becomes an aesthetic mode and an economic logic. Through case studies and open dialogue, the session interrogated the implications of this accelerated reproductive cycle—its potential for distinctive forms of expression (such as vibe musicking), and its entanglement with platform capitalism, data exploitation, and the erosion of provenance in contemporary musical culture.

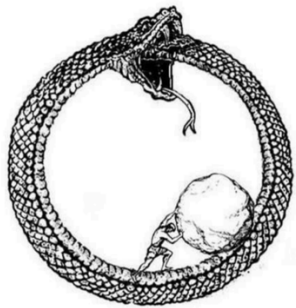

**Figure A.7.1.** Slide diagnosing platform-scale overproduction—self-replenishing content loops that exhaust meaning while normalising perpetual output.

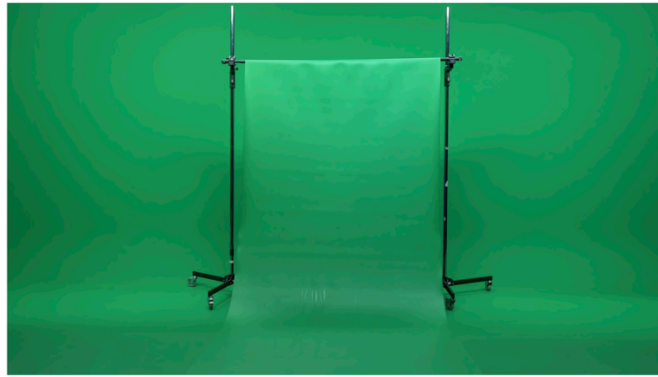
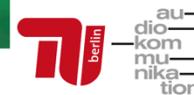

**Figure A.7.2.** Slide illustrating feed-driven hyperreality, where synthetic and recorded sources collapse into a green-screened composite, eroding provenance and simulation.

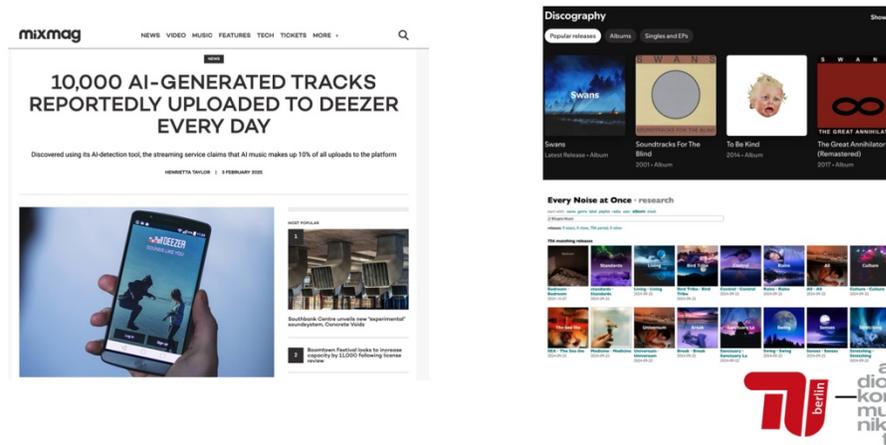
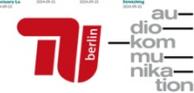

**Figure A.7.4.** Slide showcasing how platforms like Spotify and recommender infrastructures are begging to implement synthetic music/data**,** inaugurating a *post-authenticity* regime.

# Appendix B Assessment Framework

Assessment in the course was designed to balance technical engagement, creative inquiry, and critical reflection, foregrounding the relationship between theory and practice across all modalities. Students were evaluated through three interrelated components: bi-weekly assignments (30%), an in-class presentation (20%), and a final practice-based research project (50%).

Each bi-weekly assignment paired hands-on experimentation with a reflective essay, inviting students to document their technical approach while articulating the conceptual, aesthetic, and critical considerations that informed their work. Assignments required students to describe their methodology in detail—specifying tools, models, and workflows—and to situate their output within broader discussions of representation, modality, and sonic materiality introduced in the course.

The in-class presentation provided an opportunity for students to publicly share and discuss a selected work-in-progress or completed project. Emphasis was placed on clearly articulating both the creative process and the conceptual framework underpinning the work, fostering a space for collective feedback, peer exchange, and dialogical reflection.

The culmination of the course was a final practice-based research project, in which students selected a modality or conceptual theme for deeper exploration. Projects consisted of both a creative portfolio and a written reflection (1,000–4,000 words), in which students articulated their aesthetic intentions, justified their use of AI tools, and critically linked their process to theoretical frameworks encountered during the course. These projects demonstrated a high degree of autonomy and conceptual specificity, often reflecting idiosyncratic engagements with latent space, sonic transformation, or intermodal translation.

Collectively, these assessment components established a comprehensive evaluative framework that engaged students across multiple dimensions of learning. By integrating technical experimentation, aesthetic development, and theoretical reflection, the assessment structure recognized that significance in AI-mediated music creation emerges not merely from technical proficiency or established aesthetic outcomes, but through the critical articulation of intention, process, and conceptual positioning. This approach underscored the course's commitment to developing thoughtful practitioners capable of navigating the complex interplay between technological innovation, creative expression, and critical discourse in contemporary sonic arts.

# Appendix C Additional Student Projects: Creative and Conceptual Explorations

### C.1 Graphemic prompting and intersemiotic translation: Artun Otter's ElevenLabs ASCII Angry Sample Pack[16]

Artun Otter's project reframed text-to-audio as a semiotic interpreter of emojis rather than a natural-language renderer. Instead of lexical prompts, Otter fed the ElevenLabs SFX generator ASCII emoticons and text art, alongside a smaller set of affect-laden phrases, to examine how a model translates non-lexical signs—shapes, spacing, punctuation density—into sound. The study thus shifted from word semantics to graphemic morphology, asking how visual marks function as prompts and what kinds of acoustic categories they elicit.

---

[16] The student's work can be accessed via the following digital repository:
https://drive.google.com/file/d/1VLeErvE702P-_qV7rYrFO85G65lm_RuV/view?usp=sharing

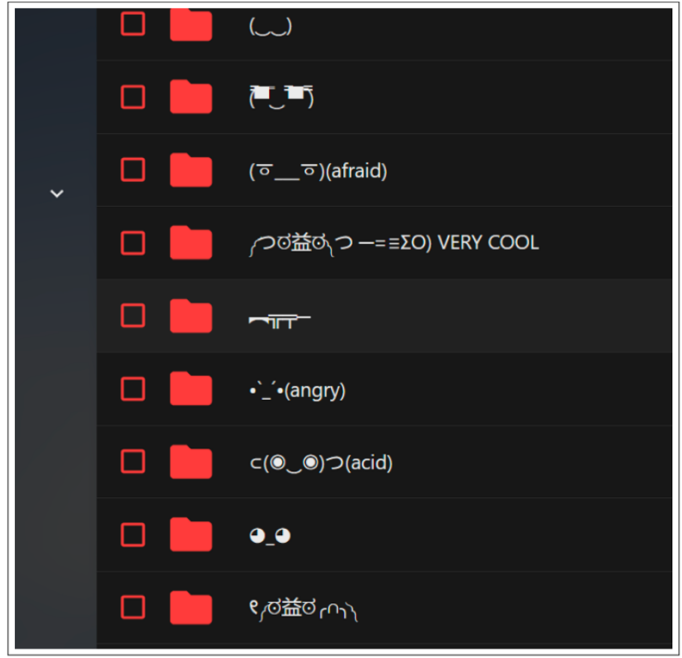

Figure . Examples of ASCII Emoticon prompts used.

Forty-three generations were produced and annotated. Prompts were grouped into two families: (1) graphemic signs (e.g., faces, "hadouken" glyphs, gun icons) whose meaning is carried primarily by form; and (2) affective phrases whose meaning is carried by lexical content. Otter compared cross-prompt consistencies, listening for category biases and recurring textures, then abstracted patterns that cut across both families.

**Findings as semiotic behaviours:**

- **Iconic coupling (form→category).** Certain ASCII shapes repeatedly mapped to acoustic classes: gun-like glyphs → gunshots; "hadouken"-style rays → explosive transients. Here, the *iconicity* of the sign (its visual resemblance or cultural shorthand) appears to bias category selection: a grapheme-to-acoustic translation rather than a word-to-sound mapping.
- **Paralinguistic voicing.** "Eyes" emoticons and similar face-like forms often yielded uncanny voice/animal vocalisations—near-speech timbres with ambiguous phonation. This suggests the model infers presence/agency from facial iconography and translates it as vocal sound, revealing a route from **pictogram** → **vocality** independent of words.
- **Affective non-linearity.** Aggressive phrases produced divergent results (from near-silence to harsh noise), indicating a non-monotonic mapping between sentiment labels and sonification. Here, symbolic meaning (lexical hostility) does not guarantee a stable acoustic correlate; the model's affective priors surface as semiotic slippage.
- **Musical bleed-through in a non-music model.** Some graphemic inputs triggered loop-like beats or jingle-adjacent idioms, evidencing latent genre priors even within an SFX-oriented system—another form of cross-code translation.

Taken together, the results show that the model performs intersemiotic translation across sign types: *iconic* (shape), *indexical* (gesture/agency implied by faces or rays), and *symbolic* (words). By moving from words to sign-forms, Otter exposed an additional semiotic channel in TTA systems—one in which typography, spacing, and glyph design act as operative features that the model

"interperts" and renders sonically. The most compelling outputs arise where this reading is unstable, producing semiotic drift (e.g., face → voice; rays → impact), which composers can treat as a structured source of material rather than error.

*ASCII Angry Sample Pack* demonstrates that TTA models can be prompted through graphemic morphology to yield consistent families of sounds, revealing a practice of graphemic prompting and affective semantics alongside conventional language prompting. In doing so, it surfaces another form of semiotic interplay and intersemiotic translation—from sign-shape to sound—and opens a compositional palette grounded in the interpretive biases of the model itself.

**C.2 Door Creak to Miles Davis: Jakob Lavric's TTA Rhizomixing Assignment[17]**

Jakob Lavric's homework pivots on a short viral clip[18] in which a street door creaks with the uncanny contour of a muted-trumpet phrase. He used this moment as a seed for exploring musical continuation in Udio: could the system recognise and extend an incidental, non-musical fragment into a coherent musical structure? The project tested how a listener's genre-conditioned imagination—here, the jazz idiom—can be externalised and amplified through text-to-audio generation.

Lavric designed a prompt scaffold—"melancholic trumpet melody," "brushed drum kit," "film noir," "sparse jazz chords"—to position the system within a coded stylistic space. Guided by this semantic frame, Udio extended the creaking "phrase" toward articulations reminiscent of *Blue in Green* and modal jazz more broadly, while maintaining the door's gestural "DNA". The prompt thus operated as semantic armature, steering the model to recognise latent musical potential in an everyday sound and to elaborate it within specific idiomatic constraints.

As Lavric articulated in his reflection:

"By extracting and extending these accidental phrases, the process materialises continuation—transforming ephemeral, unintended sounds into structured musical expression. This process not only bridges the gap between randomness and intention but also highlights how we are constantly surrounded by latent musicality in everyday life—waiting to be recognised, shaped, and set into motion." (Lavric, 2024)

In effect, the piece shows how found sonic coincidence (and any sound) can be channelled into genre-aware development through careful prompt design and TTA systems. Taken together, the study demonstrates prompt-conditioned continuation, genre-inference alignment, and the transformation of incidental environmental sound into structured musical form within contemporary TTA workflows.

---

[17] The student's work can be accessed via the following digital repository:
https://drive.google.com/drive/folders/1rwvj5uj-nkILpJyLaC6jAYmMwigoAZMc?usp=sharing

[18] The video clip that inspired this project can be viewed at:
https://www.youtube.com/watch?v=wwOipTXvNNo